# EVALUATION OF COMPANY INVESTMENT VALUE BASED ON MACHINE LEARNING

Junfeng Hu[1], Xiaosa Li[2], Yuru Xu[3], Shaowu Wu[4] and Bin Zheng[5]

[1,2,3,4,5] Faculty of Science, School of Mathematics,
Beijing University of Technology, Beijing, China,
[1]1187618301@qq.com, [2]1660223054@qq.com, [3]yuruxu123@163.com
[4]shaowu@emails.bjut.edu.cn, [5]zhengbin@bjut.edu.cn

## ABSTRACT

*In this paper, company investment value evaluation models are established based on comprehensive company information. After data mining and extracting a set of 436 feature parameters, an optimal subset of features is obtained by dimension reduction through tree-based feature selection, followed by the 5-fold cross-validation using XGBoost and LightGBM models. The results show that the Root-Mean-Square Error (RMSE) reached 3.098 and 3.059, respectively. In order to further improve the stability and generalization capability, Bayesian Ridge Regression has been used to train a stacking model based on the XGBoost and LightGBM models. The corresponding RMSE is up to 3.047. Finally, the importance of different features to the LightGBM model is analysed.*

## 1. INTRODUCTION

Company investment value assessment is an outcome of active market in combination with modern enterprise systems, which can guide investors to understand the intrinsic value of a company. It includes both theoretical analysis and practice which help investors identify valuable investment projects, and make correct and reasonable investment decisions.

At present, traditional methods for company investment value evaluation mainly include free cash flow discount method, economic added value model, price-earnings ratio method, and so on. In particular, factor analysis and analytic hierarchy process are examples of empirical evaluation methods for investment value [1, 2, 3]. With the increase of information content and the development of big data analysis techniques, effective methods have also been developed to reduce uncertainty in real-time decision-making [4, 5]. In [6], a linear information model for value evaluation has been proposed. Random forest and support vector machine are used to establish a high-precision enterprise investment evaluation system [7]. Information transparency is shown to have a significant influence on the company's investment value based on unbalanced panel random-effects regression [8]. In [9], both financial data and non-financial data are trained by machine learning to establish a comprehensive evaluation model for enterprise investment value.

In this work, based on a data set containing comprehensive company information, a variety of data mining techniques and machine learning algorithms are used to develop effective company investment value evaluation models.

## 2. DATA AND METHODS

In this section, data processing, including pre-processing and feature extraction, will be discussed. Then, a detailed description of the models used in this study will be given.

### 2.1. Data Source

Data in this paper are from IEEE ISI World Cup 2019, including industrial and commercial information, annual report, financial information, tax information, equity information, legal information, intellectual property information, business information, land purchase information and other data of 3500 listed company (a total of 37 excel sheets and 1 enterprise rating form) [10]. These data come from official statistical platform, the data is real and credible.

### 2.2. Data Pre-processing

We perform data exploratory analysis to discover the inherent data characteristics, which helps us choose appropriate techniques for data pre-processing and analysis. In particular, the quality of data set after data pre-processing is very important for feature extraction in the next step. Hence, data pre-processing has a great impact on model results. The following is a list of problems involved in the data pre-processing and the corresponding processes taken:

- Label, Sample duplication problem: Deduplication.
- Missing value filling problem: Filling 0 or -99.
- Category variable processing: Label encoding and One-hot encoding.
- A Unit conversion: 1 Dollar = 6.7*1 Yuan, 100 Million Yuan = 100,000,000 Yuan.
- Date conversion: Convert to a timestamp, Separation of year, month, day, convenient extraction of time characteristics.
- Credit rating conversion: Advanced certification enterprise = 4, General certified enterprise = 3, General credit enterprise = 2, The remaining = 1.
- Exception string handling: For example '--', '\xad'. replace with -99.
- Number extraction: Mostly Regular Expression.

### 2.3. Feature Extraction

To facilitate feature extraction, each Excel form is extracted separately. Then, all features are combined into one Excel form. The features extracted in this paper are shown in TABLE I, where the first column is the variable names in the original table, and the second column is the corresponding feature extraction.

In the TABLE 1, count is the number of occurrences of some data, mean is the average of data, nunique is the number of elements in the data set, max is the maximum value of the data, min represents the minimum value of data, std is the standard deviation of data, skew means the skew of the data, median is the median of the data. In total, there are 436 features extracted from the data set.

TABLE 1. Original variable and its features

| Original Variables | Feature Extraction |
|---|---|
| Company Number | count: The company number in each table |
| Product Type | count, nunique |
| Registered Capital (Ten Thousand Yuan), | Registered Capital (Ten Thousand |

| Original Variables | Feature Extraction |
|---|---|
| Number of Employees | Yuan), Number of Employees |
| Registered Capital Currency (Regular), Operating State, Industry Categories (Code), Industry of Small Class(Code), Type, Province Code, City Code, Area Code, Whether the Listed, Registration Authority Area Code | LabelEncoder |
| Industry Categories (Code) | OneHotEncoder |
| Cancellation Reason | 1- No missing value, 0- Missing value |
| Date of Establishment | The time interval between the date of establishment and the term of operation |
| Land Use, Administrative Region | nunique |
| Total Land Supply Area, Transaction Price (Ten Thousand Yuan) | Median、mean、max、min、std、skew |
| Total Area | mean |
| Economic Division | count: Each economic division |
| Tax Year A | count |
| Credit Rating | mean |
| Annual Report Year | Count, mean, max, std: Annual reports from 2013 to 2017 |
| Label of Competing Products, Competitive Product Rotation, Detailed Address of Competing Products, Operation Status of Competing Products | count, nunique |
| Information of Subscribed Capital Contribution, Information of Paid-in Capital Contribution | mean: Logarithm |
| Are There Any Websites or Outlets, Whether the Enterprise Has Investment Information or Purchase Equity of Other Companies, Whether There Is a Change of Shareholders' Equity in the Limited Liability Company This Year, Whether to Provide External Guarantee, Whether the Number of Employees Is Open | sum |
| Trademark | count: Each state of the trademark |
| Application Date | mean,max,std,skew,kurt: The first difference of the application date<br><br>nunique, max, min, mean, median: Year of application, Day of application, |
| Earnings Per Share, Net Assets Per Share (Yuan), Provident Fund Per Share (Yuan), Undistributed Profit Per Share (Yuan), Operating Cash Flow Per Share (Yuan) | mean, std |
| Asset-liability Ratio (%), Current Liabilities/Total Liabilities (%), Liquidity Ratio, Quick Ratio | mean, std |
| Total Revenue (Yuan) | mean |
| Total Asset Turnover (Times), Days of Receivables Turnover (Days), Inventory | mean |

| Original Variables | Feature Extraction |
|---|---|
| Turnover Days (Days) | |
| Assets: Monetary Capital (Yuan), Assets: Fixed Assets (Yuan), Assets: Intangible Assets (Yuan), Assets: Total Assets (Yuan), Liabilities: Total Liabilities (Yuan) | mean |
| Label, Provinces, Name of the Certificate | nunique |
| Coupon Rate (%), Total Planned Issuance (100 Million Yuan) | mean |
| Patent Type | count: Each type of patent type |
| State | count: Each type of state |

### 2.4. Feature Selection

When the number of features increases, it may cause "dimension disaster" and hence decrease model learning performance [11]. To avoid such problems including the over-fitting problem, many feature selection methods have been developed, such as filter, wrapper, and embedded method. Here, a gradient boosting decision-tree model is chosen to reduce feature dimensionality. This method combines the advantages of the filter and wrapper methods, and uses the parameters inside the learner to sort the features. This effectively improved the performance of the learner and the computing efficiency [12].

### 2.5. Models

This section introduces the main models used in this paper: XGBoost model, LightGBM model and Stacking model.

First, XGBoost (Extreme Gradient Boosting) is a popular gradient boosted trees algorithm. Traditional GBDT (Gradient Boosting Decision Tree) uses first-order derivative information for optimization [13]. XGBoost uses a second-order Taylor expansion of the loss function, which adds second derivative information in addition to the retained first-order derivative. Hence, it can speed up model convergence on the training set [14].

XGBoost allows for setting sample weights. By adjusting these weights, we can pay more attention to some samples and use many strategies to prevent overfitting, such as introducing regularization term, Shrinkage and Column Subsampling, etc. Support of parallelism is one of the great strengths of XGBoost, which allows nodes of the same hierarchy to run in parallel. XGBoost was also designed to handle sparse data effectively.

Second, LightGBM (Light Gradient Boosting Machine) was released by Microsoft Asia research Institute in 2016. It is an open source, fast and efficient promotion framework based on decision tree algorithm. It is used in sort, classification and regression, and other machine learning tasks, and supports efficient parallel training [15]. LightGBM contains gradient-based one-side Sampling (GOSS) and Exclusive Feature Bundling (EFB). GOSS is used to filter partial data. The role of EFB is to bind and merge mutually exclusive features, thus achieving dimension reduction. The so-called mutual exclusion means that there are some features in the feature space that do not take non-zero values at the same time [16, 17]. LightGBM uses the leaf-wise method with depth limitation to improve the accuracy of the model. By changing the decision rules of the decision tree algorithm, LightGBM provides direct native support for categorical features without transformation, hence greatly improves the training speed.

Third, stacking algorithm is an integrated learning algorithm proposed by Wolpert in 1992 [18]. For Bagging and Boosting method the same learning algorithm is used in a single training study [19, 20]. Stacking algorithm combines a number of different learning algorithms to improve generalization performance. It can be considered as a special kind of portfolio strategy, similar to the Voting and Blending [21]. Moreover, the stacking algorithm is constructed by cross validation and hence robust.

## 3. Results

This work first applies XGBoost and LightGBM for 5-fold cross-validation, and then use stacking algorithm for model fusion to improve stability and generalization ability. The performance of these models on feature set with or without feature selection is reported. Here, Root-Mean-Square Error (RMSE) is selected to measure model performance. The corresponding formula is given in Equation 1.

$$RMSE = \sqrt{\frac{\sum_{i=1}^{n}(X_{obs,i} - X_{model,i})^2}{n}} \qquad (1)$$

where $X_{obs,i}$ is the true value, $X_{model,i}$ is the predicted value, and $n$ is the sample number.

The results of models performance are shown in TABLE 2. It can be seen from TABLE 2 that LightGBM model is superior to XGBoost model when using the same feature set. Model fusion can achieve higher accuracy. Also, the accuracy is improved after feature selection. Hence, more features does not necessarily lead to higher accuracy. A better selected feature set may help to achieve better model performance.

TABLE 2. Model Results

| **Models** | **RMSE** | |
| --- | --- | --- |
| | *No feature selection, number of features is 436* | *With feature selection, number of features is 66* |
| XGBoost | 3.109 | 3.098 |
| LightGBM | 3.065 | 3.059 |
| Stacking fusion | 3.056 | 3.047 |

The importance of features for LightGBM is ranked, and the top 10 are shown in Fig. 1. Registered Capital is the most important, followed by the standard deviation of Current Liabilities/Total Liabilities, the mean of Total Asset Turnover, the standard deviation of Earnings Per Share and the mean of Provident Fund Per Share. The importance of features can help investors quickly select features they are interested in and make decisions based on various factors.

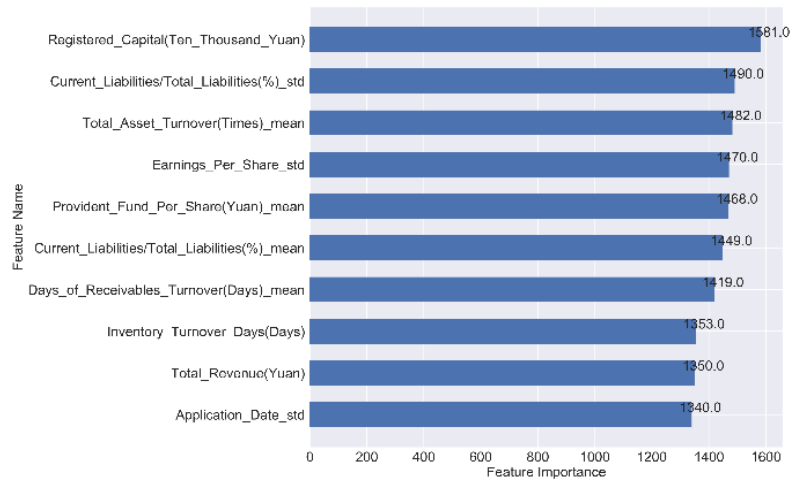

Fig 1. The importance of features for LightGBM

## 4. CONCLUSIONS

In this paper, several company investment value evaluation models are proposed by mining the comprehensive company information, extracting valuable features, and applying machine learning algorithms. The stacking model can achieve high precision. The obtained features importance value from LightGBM is instrumental for company investors. In order to find more valuable features and improve the accuracy of the proposed models, advanced feature engineering techniques as well as deep learning algorithms will be explored in future work.

## ACKNOWLEDGEMENTS

The research in this paper has been completed successfully, thanks to the platform provided by IEEE International Conference on ISI and Shenzhen Artificial Intelligence and Data Science Research Institute, as well as the data support.